\documentclass[9pt,letterpaper]{article}

\usepackage[margin=1in]{geometry}
\usepackage[T1]{fontenc} 
\usepackage{graphicx}

\usepackage{xcolor}
\usepackage{amsmath}

\usepackage{footmisc}
\usepackage[parse-numbers=false]{siunitx}
\usepackage{amsfonts}
\usepackage{upgreek}
\usepackage{subcaption}
\usepackage{bm}
\usepackage{placeins}

\usepackage{algorithm}
\usepackage{algpseudocode}
\usepackage[resetlabels]{multibib}
\newcites{app}{References}

\usepackage{cite}

\usepackage{booktabs}
\usepackage[font=small,labelfont=bf]{caption}

\usepackage{float}

\DeclareMathOperator*{\minimize}{minimize}
\DeclareMathOperator{\sign}{sign}

\definecolor{dkgreen}{rgb}{0,0.6,0}
\definecolor{dkred}{rgb}{0.6,0,0}
\definecolor{dkblue}{rgb}{0,0,0.6}

\title{\Huge\textbf{Data-driven control of a\\ magnetohydrodynamic flow}\thanks{This work was co-funded by the Ministry of Education, Youth and Sports of the Czech Republic, and European Commission within the project ROBOPROX (reg. no. CZ.02.01.01/00/22\_008/0004590).}}
\author{Adam Uchytil$^{1,}$\thanks{Corresponding author. Email address: \texttt{uchytada[at]fel.cvut.cz}} \and Milan Korda$^{1,2}$ \and Jiří Zemánek$^{1}$}

\date{\today}

\begin{document}

\footnotetext[1]{Czech Technical University in Prague, Faculty of Electrical Engineering, 166 27 Prague, Czechia.}
\footnotetext[2]{LAAS-CNRS, Université de Toulouse, 31400 Toulouse, France.}

\maketitle

\begin{abstract}
We demonstrate the feedback control of a weakly conducting magnetohydrodynamic (MHD) flow via Lorentz forces generated by externally applied electric and magnetic fields.
Specifically, we steer the flow of an electrolyte toward prescribed velocity or vorticity patterns using arrays of electrodes and electromagnets positioned around and beneath a fluid reservoir, with feedback provided by planar particle image velocimetry (PIV).
Control is implemented using a model predictive control (MPC) framework, in which control signals are computed by minimizing a cost function over the predicted evolution of the flow.
The predictor is constructed entirely from data using Koopman operator theory, which enables a linear representation of the underlying nonlinear fluid dynamics.
This linearity allows the MPC problem to be solved by alternating between two small and efficiently solvable convex quadratic programs (QPs): one for the electrodes and one for the electromagnets.
The resulting controller runs in a closed loop on a standard laptop, enabling real-time control of the flow.
We demonstrate the functionality of the approach through experiments in which the flow is shaped to match a range of reference velocity fields and a time-varying vorticity field.
\end{abstract}


\section*{Significance}
Contactless actuation of fluid flows is highly relevant to various applications, such as mixing, cooling, and chemical processing.
Conducting fluids can utilize this unique actuation opportunity via electromagnetic fields. 
However, controlling these flows is challenging due to the inherent complexity of fluid dynamics and its computational demands. 
We present one of the first experimental demonstrations of closed-loop, data-driven control of a conducting fluid via electromagnetic fields. 
We shape an electrolyte flow into desired velocity and vorticity patterns using electromagnets, electrodes, and visual feedback. 
Leveraging Koopman operator theory, a machine learning framework, we construct a lightweight controller entirely from flow measurements, achieving millisecond-scale computation times. 
Our method generalizes beyond fluid systems, providing a practical, model-free framework for controlling complex physical systems.

\section*{Introduction}
Controlling fluid flows is inherently challenging due to their nonlinear, high-dimensional, and often chaotic nature. 
Fluid dynamics is notoriously difficult to model in a way that balances accuracy with computational efficiency.
This challenge is particularly critical for real-time, model-based control, which often must run on embedded systems with limited computational resources and high sampling rates.
When fluid dynamics---governed by the Navier--Stokes equations---are coupled with electromagnetic interactions described by Maxwell's equations, another layer of complexity is introduced to the problem.
These flows, referred to as magnetohydrodynamic (MHD) flows, play a crucial role in numerous applications across engineering and science, including mixing and pumping \cite{bauMinuteMagnetoHydro2001, westApplicationMagnetohydrodynamicActuation2002a, qianMagnetohydrodynamicStirrerStationary2005, homsyHighCurrentDensity2005, al-habahbehReviewMagnetohydrodynamicPump2016, matiaMagnetohydrodynamicLevitationHighperformance2022, bauApplicationsMagnetoElectrochemistry2022}, astrophysical and geophysical phenomena \cite{davidsonIntroductionMagnetohydrodynamics2016}, and heat transfer systems \cite{smolentsevMHDThermofluidIssues2010,oharaMagnetohydrodynamicbasedInternalCooling2023}.
A distinctive feature of MHD flows is that they can be influenced without mechanical components, enabling fully contactless actuation of the fluid.

One way to obtain tractable models of fluid flows is through analytical model order reduction techniques, which project the original high-dimensional dynamics onto a lower-dimensional subspace.
These techniques have been successfully applied to MHD flows in conjunction with control, particularly in the context of simulated one- and two-dimensional channel flows \cite{ravindranRealTimeComputationalAlgorithm2005, renDynamicOptimalControl2018, renOptimalTrackingControl2019, chenRealtimeComputationalOptimal2020}.
While effective for simple flows, these methods struggle with highly nonlinear or turbulent systems, where the dynamics cannot be confined to a low-dimensional subspace.
Furthermore, the resulting reduced-order models often require nonlinear control strategies, which are computationally expensive, or rely on local linearization and linear control design, which limits the global applicability of the control strategy.

Recently, machine learning (ML) has emerged as a powerful tool across many areas of fluid mechanics, including modeling, simulation, optimization, and control \cite{bruntonMachineLearningFluid2020,vinuesaTransformativePotentialMachine2023a}.
In particular, deep learning (DL) methods have been successfully applied to model complex fluid flows to enhance simulations \cite{vinuesaEnhancingComputationalFluid2022}, and provide closure models for turbulence \cite{sirignanoDeepLearningClosure2023}.
These DL-based models have shown promise in accurately capturing the dynamics of fluid flows from data without the knowledge of the governing equations.
When combined with model predictive control (MPC), such models have been effectively used to control simulated fluid flows \cite{mortonDeepDynamicalModeling2018, biekerDeepModelPredictive2020}. Deep reinforcement learning has also emerged as an alternative paradigm for learning control policies directly from interaction with the flow \cite{garnierReviewDeepReinforcement2021,viqueratReviewDeepReinforcement2022, vignonRecentAdvancesApplying2023}.
However, DL models have their drawbacks, such as large data requirements, difficult interpretability, and nonlinearity, leading to high computational costs that may hinder their application in real-time control.

Koopman operator theory (KOT) \cite{koopmanHamiltonianSystemsTransformation1931, koopmanDynamicalSystemsContinuous1932, mezicComparisonSystemsComplex2004, mezicSpectralPropertiesDynamical2005, budisicAppliedKoopmanism2012} provides an alternative ML framework for data-driven modeling of dynamical systems.
KOT represents the dynamics of a nonlinear system as a linear operator acting on a space of observables---known as the Koopman operator. 
Although the Koopman operator is generally infinite-dimensional, data-driven methods such as dynamic mode decomposition (DMD) \cite{schmidDynamicModeDecomposition2010,williamsDataDrivenApproximation2015, proctorDynamicModeDecomposition2016} can be used to construct finite-dimensional approximations.
These approximations serve as global linear representation of the underlying nonlinear dynamics, enabling the use of efficient control strategies designed for linear systems, such as linear quadratic regulators or linear model predictive control  \cite{ottoKoopmanOperatorsEstimation2021,bevandaKoopmanOperatorDynamical2021}.
Compared to DL, KOT-based methods are more data-efficient and provide greater interpretability. 
In fluid mechanics, KOT has been successfully applied for both analysis \cite{rowleySpectralAnalysisNonlinear2009, mezicAnalysisFluidFlows2013, arbabiStudyDynamicsPosttransient2017} and control \cite{arbabiDataDrivenKoopmanModel2018a, peitzKoopmanOperatorbasedModel2019b, peitzUniversalTransformationDatadriven2023}.

In this paper, we introduce a systematic approach to real-time control of MHD flows actuated by Lorentz forces generated by external electric and magnetic fields.
Our approach is fully data-driven and enabled by Koopman model predictive control (KMPC) \cite{kordaLinearPredictorsNonlinear2018}.
We experimentally validate our control strategy using a novel laboratory setup that employs arrays of individually controllable electrodes and electromagnets to generate both electric and magnetic fields---a key distinction from previous platforms, which only controlled the electric field \cite{henochExperimentalInvestigationSalt1995, breuerActuationControlTurbulent2004, pangTurbulentDragReduction2004}---and uses real-time planar velocity measurements to provide feedback to the controller.
Importantly, the controlled flow is inherently three-dimensional and cannot be reduced to a two-dimensional approximation, which distinguishes our system from many simplified benchmark settings (e.g. Hele--Shaw cells \cite{mckeeMagnetohydrodynamicFlowControl2024}).
Through our approach and experimental setup, we present what we believe to be among the first experimental realizations of data-driven real-time fluid flow control in a closed-loop setting with optimization-based actuation, paving the way for more sophisticated applications in fluid mechanics and beyond.

\section*{Results}
We address the problem of controlling the flow of a weakly conducting fluid.
Specifically, we aim to drive the flow toward a prescribed reference velocity or vorticity field, representing structures such as vortices, jets, or shear layers, depending on the intended application.
This is achieved using external electric and magnetic fields, $\bm{E}$ and $\bm{B}$, which interact to produce a Lorentz force that drives the flow.
Assuming a low magnetic Reynolds number and negligible motional electromotive force, the Lorentz force density can be written as
\begin{equation}
    \bm{f} = \sigma \bm{E} \times \bm{B}, \label{eq:lorentz_force}
\end{equation}
where $\sigma$ is the fluid's electrical conductivity (see SI~Appendix~\ref{app:lorentz_force} for a derivation and justification).
The fields are generated by superposing contributions from individually controllable actuators---$n_\text{el}$ electrodes producing electric fields and $n_\text{mag}$ electromagnets generating magnetic fields.
For control design purposes, we consider the fields as a linear combination of time-independent spatial basis fields modulated by time-dependent control signals
\begin{equation}
    \bm{E}(t,\bm{r}) = \sum_{i=1}^{n_\text{el}} \phi_i(t) \bm{E}_i(\bm{r}), \quad \bm{B}(t,\bm{r}) = \sum_{j=1}^{n_\text{mag}} \psi_j(t) \bm{B}_j(\bm{r}), \label{eq:fields_decomposition}
\end{equation}
where $\phi_i(t)$ and $\psi_j(t)$ are the control signals, and $\bm{E}_i(\bm{r})$, $\bm{B}_j(\bm{r})$ are spatial basis fields associated with each actuator.
This decomposition induces a bilinear structure in the Lorentz force density 
\begin{equation}
    \bm{f}(t, \bm{r}) =  \sigma \sum_{i=1}^{n_\text{el}} \sum_{j=1}^{n_\text{mag}}\phi_i(t) \psi_j(t) \;\bm{E}_i(\bm{r}) \times \bm{B}_j(\bm{r}). \label{eq:lorentz_force_expanded}
\end{equation}
By adjusting the vectors of control signals $\bm{\phi}(t) = \left[\phi_1(t), \ldots, \phi_{n_\text{el}}(t) \right]^\top$ and $\bm{\psi}(t) = \left[\psi_1(t), \ldots, \psi_{n_\text{mag}}(t) \right]^\top$, we directly shape the Lorentz force and thereby influence the fluid flow.
For implementation, these dimensionless control signals are mapped to physical actuator commands through system-specific transformations (see SI~Appendix~\ref{app:fields_decomposition} for specifics).

To compute the control signals corresponding to a given reference flow field, we adopt a feedback control approach.  
This requires real-time velocity measurements within the controlled region of the flow, which are used for continuous adjustment of the control inputs.
We next describe the control methodology in detail, followed by an overview of the experimental setup and the validation experiments conducted. 

\subsection*{Koopman model predictive control}
\begin{figure*}[!t]
    \includegraphics{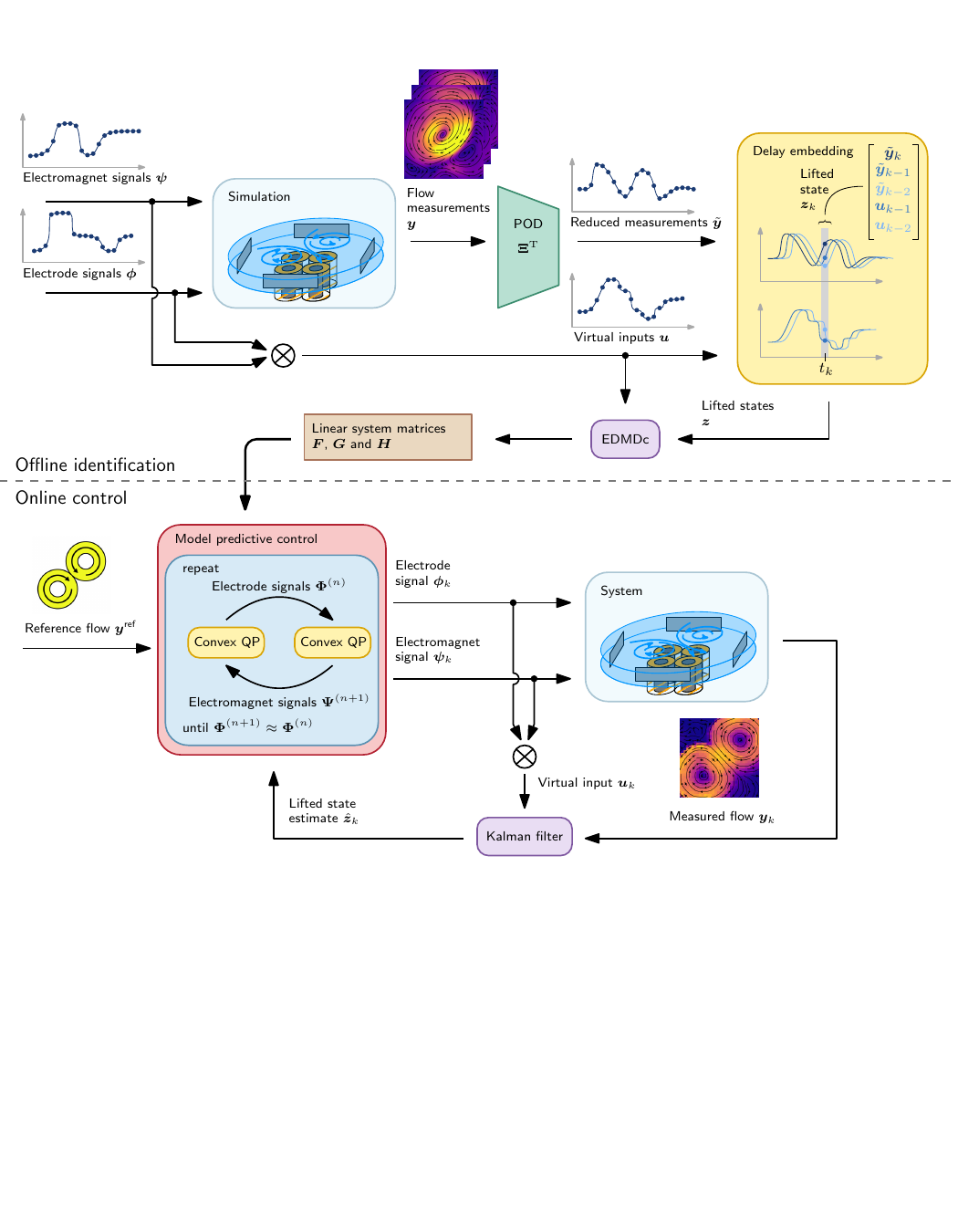}
\caption{
Overview of our framework for real-time control of MHD flows. \textbf{Top:} In the offline phase, actuation signals from electrodes and electromagnets generate flow data, which is reduced via POD, delay-embedded, and used to learn a linear predictor via EDMDc. \textbf{Bottom:} In the online phase, a KF estimates the system state from measurements, and the controller solves two alternating convex QPs to compute real-time actuation signals that steer the flow toward a target velocity field. The virtual input structure $\bm{u} = \bm{\psi} \otimes \bm{\phi}$ captures the bilinear nature of the Lorentz force.
}
\label{fig:full_diagram}
\end{figure*}

We begin by describing our approach to data-driven control of fluid flow using the Koopman operator, which we also illustrate in Figure~\ref{fig:full_diagram}.
We use a a Koopman-based model of the fluid dynamics to formulate an optimization problem that minimizes the discrepancy between predicted and desired velocity fields over a finite time horizon. 
Solving this problem yields a sequence of control signals for the electrodes and electromagnets over the prediction horizon.
The optimization is initialized using current velocity field measurements, and only the first signal from the optimal sequence is applied to the actuators.
As new measurements become available, the process is repeated, forming a closed-loop control system. This type of controller is known as Koopman model predictive control (KMPC) \cite{kordaLinearPredictorsNonlinear2018}.

We learn the Koopman-based model from a dataset consisting of velocity field measurements of the fluid flow, and the corresponding actuator signals by which they were generated.
In our case, we use data from high-fidelity simulations of the flow but the approach applies equally to experimental measurements.
The model is represented as a discrete-time linear time-invariant system parameterized by matrices $\bm{H}$, $\bm{F}$ and $\bm{G}$, serving as a finite-dimensional approximation of the Koopman operator.
The Koopman operator is a linear operator that evolves functions of the fluid's velocity---referred to as observables---forward in time by a fixed time step $\Delta t$.

To approximate the Koopman operator, we use extended dynamic mode decomposition with control (EDMDc) \cite{williamsDataDrivenApproximation2015, kordaLinearPredictorsNonlinear2018}.
We define the observables as a concatenation of delayed velocity field measurements and past actuator signals---a technique known as delay-coordinate embedding.
Prior to delay embedding, we project the velocity fields onto a low-dimensional basis using proper orthogonal decomposition (POD) to reduce dimensionality and improve numerical conditioning.
To account for the bilinear nature of the Lorentz force, we do not use the electrode and electromagnet signals directly. Instead, we define a virtual input $\bm{u} = \bm{\psi} \otimes \bm{\phi}$, where $\otimes$ denotes the Kronecker product.
The EDMDc algorithm then fits the model by expanding the dataset in the basis of these observables and solving a least-squares problem to determine the matrices $\bm{F}$ and $\bm{G}$, while $\bm{H}$ is determined by the POD modes.
This yields a closed-form solution that can be computed in seconds on a standard laptop.

Solving the optimization problem within the controller during the short time allotted by real-time constraints is challenging.
There are two main contributing factors:
First, the Koopman-based model is high-dimensional, requiring a large number of observables to accurately capture the complexity of the fluid flow, resulting in a large-scale optimization problem. 
Second, the multiplication of actuator signals in the virtual input $\bm{u}$ renders the optimization problem non-convex, making it inherently harder to solve efficiently.

To address the issue of dimensionality, we exploit the linearity of the Koopman operator-based model, which allows us to reduce the optimization problem to a formulation where only the actuator signals are treated as decision variables.
This results in an optimization problem of size $N_\mathrm{p}(n_\text{el} + n_\text{mag})$, where $N_\mathrm{p}$ denotes the length of the prediction horizon, that is, the number of discrete time steps of size $\Delta t$ over which the controller predicts the flow evolution.
This approach is commonly referred to as the dense (also sequential) MPC formulation within the control community.

To resolve the non-convexity, we leverage the structure of the virtual input $\bm{u}$---and the Lorentz force \eqref{eq:lorentz_force}---which enables the optimization problem to be decomposed into two alternating convex subproblems: one for the electrode signals and one for the electromagnet signals.
Each subproblem takes the form of a small convex quadratic program (QP) of size $N_\mathrm{p}n_\mathrm{el}$ or $N_\mathrm{p}n_\mathrm{mag}$ and can be solved in a few hundred microseconds on a standard laptop---fast enough for real-time control. 
This decomposition is a form of two-block coordinate descent algorithm (see e.g. \cite{bertsekasNonlinearProgramming2016a}), and we discuss its convergence properties in SI~Appendix~\ref{app:convergence}.

To improve robustness against measurement noise, we estimate the internal state of the Koopman model using a linear Kalman filter, which fuses velocity measurements with control inputs to produce smoothed state estimates. 
This approach eliminates the need to reconstruct delay embeddings directly from noisy data at runtime and was found to significantly improve closed-loop performance in practice.

\subsection*{Experimental setup}
\begin{figure}[tbhp]
    \centering
    \includegraphics{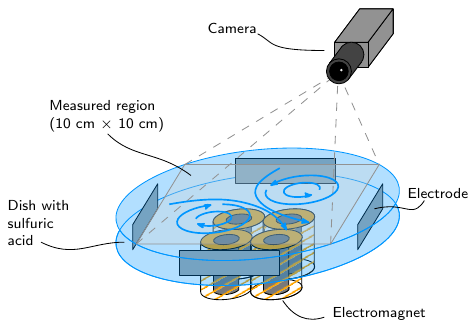}
    \caption{
Schematic of the experimental setup used for flow control. A shallow dish filled with a electrolyte (sulfuric acid) is actuated by electrodes placed along the rim and solenoidal coils positioned beneath. A top-mounted camera records tracer particle motion within a 10~cm~$\times$~10~cm region, enabling planar velocity reconstruction via PIV.
}
    \label{fig:platform}
\end{figure}

We demonstrate the functionality of our approach using a laboratory setup consisting of a dish filled with a water-based electrolyte.
Four electrodes are placed around the periphery of the dish, and four solenoidal electromagnets are positioned underneath it.
For scale, the dish has a diameter of \SI{143}{\milli\metre}, and the fluid depth is \SI{8}{\milli\metre}. 
The surface of the fluid is seeded with tracer particles, whose motion is captured by a camera providing a two-dimensional top-down view of the fluid flow within an area of \SI{10}{\centi\metre} $\times$ \SI{10}{\centi\metre} region, that is, only a portion of the dish's surface is observed.
By applying a particle image velocimetry (PIV) algorithm to the captured images, we reconstruct the planar velocity field of the fluid flow, which serves as feedback for the controller.
We illustrate this setup in Figure \ref{fig:platform}.

\subsection*{Velocity control}
\begin{figure*}[!t] 
    \includegraphics{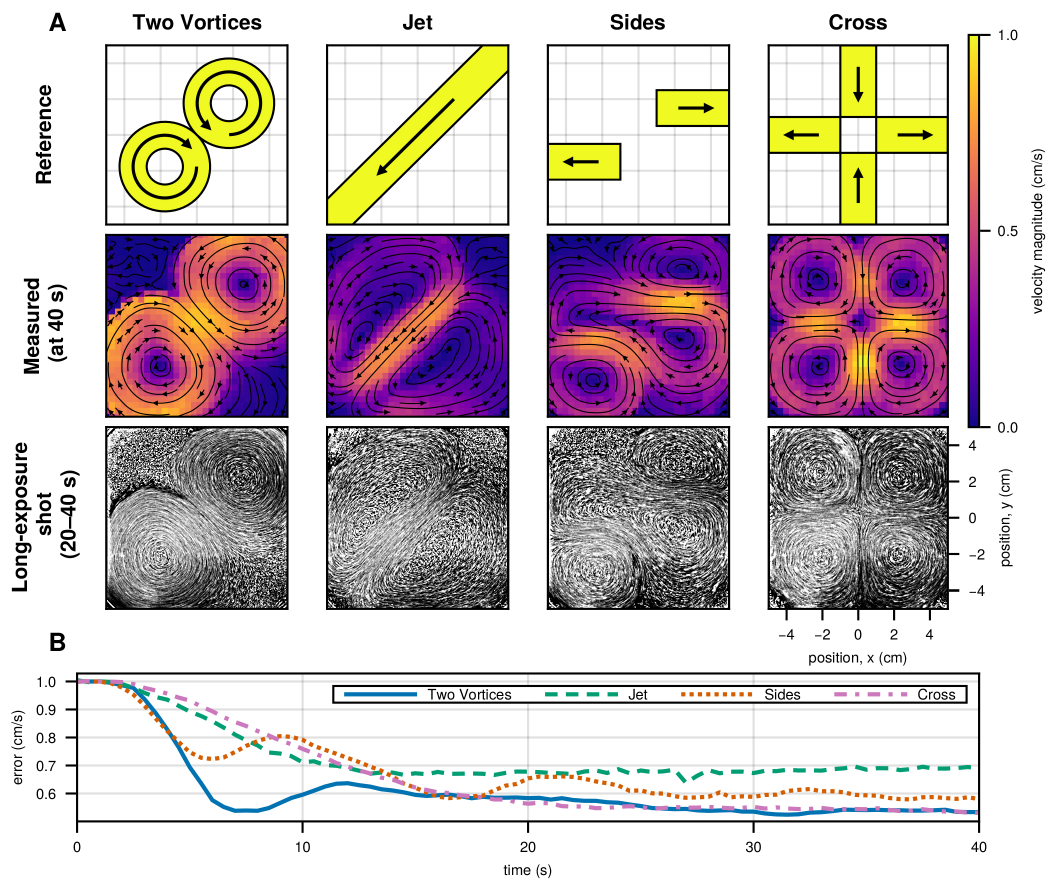}
    \caption{
Velocity field control results. 
(\textbf{A}) Target velocity fields (top), measured velocity fields at the end of each experiment (middle), and long-exposure visualizations of the flow over the final 20~s (bottom), shown for four reference patterns: \emph{Two Vortices}, \emph{Jet}, \emph{Sides}, and \emph{Cross}. 
Long-exposure images were computed by exponentially averaging histogram-equalized camera frames, while a uniform sharpening filter was applied to the resulting images to enhance visibility of the flow patterns.
(\textbf{B}) Time evolution of the velocity tracking error for each target field.
}
    \label{fig:experiments}
\end{figure*}

To demonstrate the capabilities of our approach, we present experiments targeting four qualitatively different reference velocity fields.
In each experiment, we prescribed a distinct reference velocity field with the uniform magnitude of \SI{1}{\centi\metre\per\second} within the measured region of the dish.
The controller's goal was to shape the fluid flow to match the reference field as closely as possible by controlling the electrodes and electromagnets.
We refer to these reference velocity fields as \emph{Two Vortices}, \emph{Jet}, \emph{Sides}, and \emph{Cross}, based on characteristic flow patterns they exhibit.
Each experiment lasted \SI{40}{\second}, and we present the results in Figure~\ref{fig:experiments}.
For each experiment, we show the reference velocity field, the measured velocity field at the end of the experiment, and a computational long-exposure image of the fluid motion integrated over the second half of the experiment.
The reference fields are prescribed only within a subset of the measured region (colored areas in Figure~\ref{fig:experiments}A).
Additionally, Figure~\ref{fig:experiments}B shows the evolution of the RMSE error between the measured and the reference velocity fields.
Lastly, we show these experiments in the accompanying video.

In all experiments, we observe the error (Figure~\ref{fig:experiments}B) decrease to a steady-state value between \SI{0.5}{\centi\metre\per\second} and \SI{0.7}{\centi\metre\per\second}, which remains nonzero because the desired velocity fields are not physically realizable.
The highest error of $\SI{0.7}{\centi\metre\per\second}$ is observed in the jet case, due to boundary confinement within the dish that prevents the formation of a true jet.
However, the measured fields (Figure~\ref{fig:experiments}A) resemble the desired fields well, demonstrating the effectiveness of our approach in shaping fluid flow.

\subsection*{Vorticity control}
\begin{figure*}[!t]
    \includegraphics{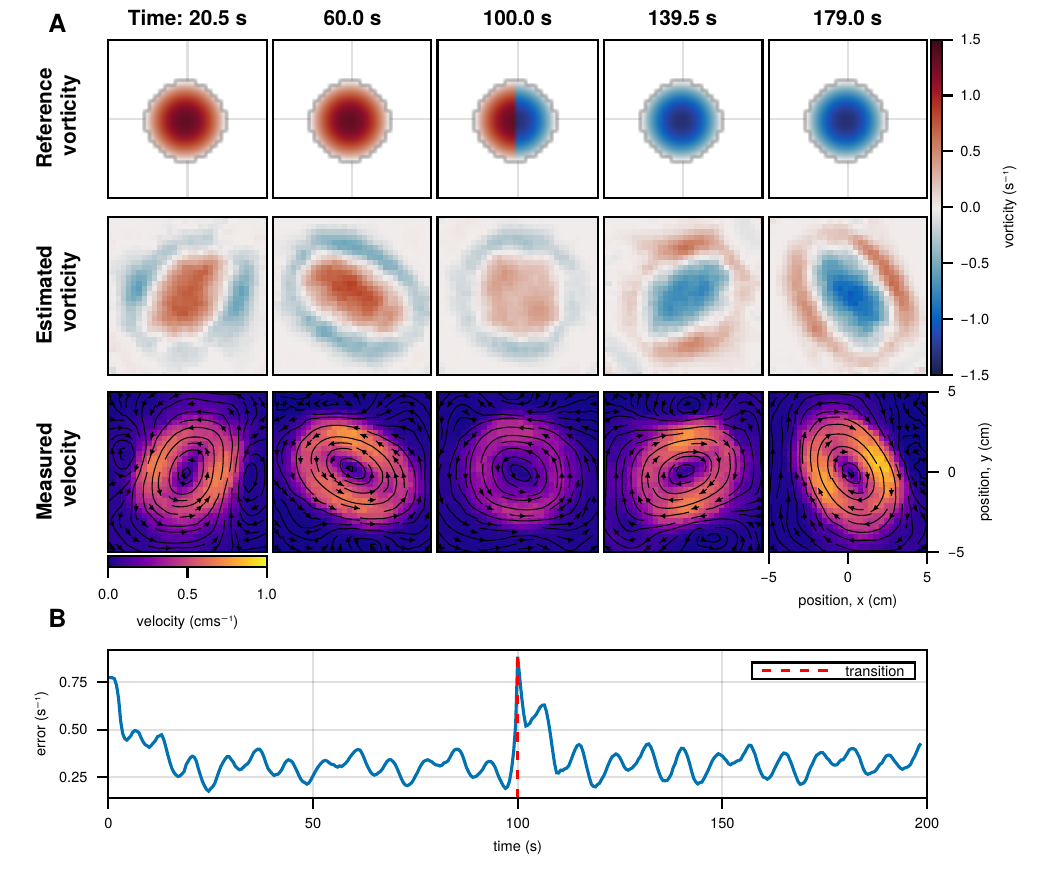}
    \caption{
Vorticity field control results. 
(\textbf{A}) Reference vorticity fields (top), estimated vorticity (middle), and measured velocity fields (bottom) at selected times. 
(\textbf{B}) Time evolution of the vorticity tracking error. A transition in the target field occurs at $t = 100$~s (red dashed line).
}
       \label{fig:vorticity}
\end{figure*}

In addition to controlling the fluid's velocity, our approach can also control its vorticity, specifically the component perpendicular to the fluid surface.
This is achieved by estimating the vorticity field from the velocity field predicted by our Koopman operator-based model and incorporating it into the cost function of the optimization problem.
We demonstrate this capability in an additional experiment, in which we prescribe a quadratic bump reference vorticity field $\omega_\text{ref} = \Omega_0 \left[ 1 - (x^2+y^2)/L^2 \right]$ for $x^2 + y^2 \leq L^2$, with $\Omega_0 = \SI{4/3}{\per\second}$ and $L = \SI{2.5}{\centi\metre}$, to generate a single vortex.
The experiment lasted $\SI{200}{\second}$, and at the $\SI{100}{\second}$ mark, we reversed the polarity of the reference field to generate a vortex with opposite rotation---demonstrating the controller's ability to track time-varying reference fields.
The results of this experiment are shown in Figure~\ref{fig:vorticity}, where we present the prescribed vorticity fields, the evolution of the RMSE between the estimated and reference vorticity fields, and snapshots of the velocity and vorticity fields throughout the experiment.

Unlike in the velocity control experiments, the error in the vorticity control case does not converge to a steady-state value but instead oscillates around $\SI{0.3}{\per\second}$.
The nonzero error is again due to the unphysical nature of the prescribed reference fields.
The oscillations are caused by the periodic evolution of the vortex, as expected, since it is not possible to maintain a single stationary vortex in our experimental setup.

\subsection*{Computational performance}
Across all closed-loop experiments, including both velocity and vorticity control, we used a prediction horizon of $N_\mathrm{p} = 10$, corresponding to predicted time of $\SI{5}{\second}$.
With $n_\text{el} = n_\text{mag} = 4$ electrodes and electromagnets, the two alternating QPs each had 40 decision variables.
The alternating optimization scheme converged in fewer than 10 full iterations on average, where each iteration consists of solving both QPs once.
Each QP took approximately \SI{300}{\micro\second} to solve on a MacBook Pro with an Apple M2 Max chip and 32 GB of RAM via the OSQP solver \cite{stellatoOSQPOperatorSplitting2020a}, resulting in a total control computation time of approximately \SI{5}{\milli\second}.
This was a small fraction of the total control loop duration, which, including PIV processing and Kalman filtering, ran at a fixed interval of \SI{0.5}{\second} without overruns.

\section*{Discussion and Conclusion}
We present a fully data-driven approach to control weakly conducting fluid flows using externally applied electric and magnetic fields.
Our method is enabled by Koopman operator theory, which allows us to construct a linear approximation of the nonlinear fluid dynamics purely from data. 
We employ this linear model within the MPC framework to synthesize an online, optimization-based controller capable of shaping the flow to match desired velocity and vorticity patterns.

Controlling both electric and magnetic fields offers more degrees of freedom in shaping the flow compared to using either field alone, but it also introduces additional complexity: the resulting Lorentz force depends bilinearly on both control inputs, rendering the MPC problem non-convex.
By exploiting both the linear structure of the Koopman model and the separable form of the Lorentz force, we reformulate the problem as two small, alternating convex quadratic programs---one for the electric field and one for the magnetic field---which can be solved efficiently on standard hardware.
This decomposition is essential for achieving real-time control and may be transferable to other systems with multiplicative control inputs.

We experimentally validate our approach on a custom-built platform consisting of a shallow dish filled with a water-based electrolyte, actuated by electrodes and electromagnets, and monitored using a planar PIV system.
Across a range of experiments, the controller reliably shaped both velocity and vorticity fields to qualitatively match the prescribed targets. 
Residual steady-state errors arise primarily because the reference fields are not physically realizable steady-state flows under our actuation constraints. 
Constructing physically feasible reference fields would require solving a coupled inverse problem involving the Navier--Stokes and Maxwell's equations---an approach that is computationally intensive, and model-dependent, and thus incompatible with the model-free nature of our framework.
Instead, we opt for intuitive, structured reference fields (e.g., vortices or jets) that represent desirable flow patterns, even if they cannot be exactly realized by the system. 
In the vorticity control experiment, oscillations in the tracking error reflect the unsteady dynamics of isolated vortices, especially under the confinement imposed by the dish geometry.

A notable strength of our framework lies in its data efficiency and real-time feasibility. 
Koopman learning requires only a single pseudo-inverse computation, and yields a globally valid linear predictor, which enables MPC.
Combined with a linear Kalman filter for state estimation, it delivers reliable closed-loop performance in the presence of measurement noise.

Despite these strengths, our current implementation has limitations. 
It relies on full-field 2D PIV measurements, which may be impractical in certain applications. 
However, this limitation could be addressed by combining sparse measurements with higher-order delay-coordinate embeddings, which have been shown effective for capturing fluid dynamics from limited sensor data \cite{arbabiDataDrivenKoopmanModel2018a}.

Overall, this work provides a blueprint for combining operator-theoretic learning with structured real-time optimization to control complex physical systems. 
By demonstrating the methodology on an experimental MHD setup, we highlight the practical potential of Koopman-based predictive control and suggest its broader applicability beyond fluid systems.

\section*{Materials and Methods}
\subsection*{Koopman operator}
We begin by briefly introducing the Koopman operator. Consider a discrete-time, possibly nonlinear dynamical system described by
\begin{equation}
    \bm{x}^{+} = \bm{S}(\bm{x}),
\end{equation}
where $\bm{x} \in M $ is the system state, $M$ is the state space, $\bm{S}$ is the state-transition map, and $\bm{x}^{+}$ is the state at the next time step, i.e., at time $t + \Delta t$.
The Koopman operator $\mathcal{K}$ is a linear operator acting on the space of observables---functions $g: M \to \mathbb{C}$.
Its action is defined as $\left(\mathcal{K} g\right)(\bm{x}) = g \left( \bm{T} (\bm{x} )\right)$, meaning that it propagates observables forward in time.
A suitable generalization of the Koopman operator for a system with control inputs was introduced in \cite{kordaLinearPredictorsNonlinear2018}.

Importantly, the Koopman operator always exists and is linear, even if the underlying system dynamics is nonlinear.
This property allows for the application of efficient and well-established control strategies designed for linear systems---such as linear MPC---to otherwise nonlinear dynamical systems.

\subsection*{Approximating the Koopman operator: Extended Dynamic Mode Decomposition with control}
The Koopman operator is typically infinite-dimensional, making it infeasible to compute directly. Therefore, data-driven methods are used to approximate it using a finite dictionary of observables.
Given such a dictionary $g_1, g_2, \ldots, g_n$, we define the lifted state as $\bm{z}(\bm{x}) = [g_1(\bm{x}), g_2(\bm{x}), \ldots, g_n(\bm{x})]^\top$, which allows us to represent the original dynamics by the linear system
\begin{subequations}
    \begin{align}
        \bm{z}^{+} &= \bm{F} \bm{z} + \bm{G} \bm{u},\\
        \hat{\bm{y}} &= \bm{H} \bm{z},
    \end{align} \label{eq:surrogate_model}
\end{subequations}
where $\bm{F}$ is the state-transition matrix, $\bm{G}$ is the input matrix, $\bm{H}$ is the output matrix, $\bm{u}$ is the control input, and $\hat{\bm{y}}$ is the predicted output of the system---in our case, the predicted velocity field.

The predictive accuracy of model \eqref{eq:surrogate_model} depends critically on the choice of observables. 
In our case, we employ delay-coordinate embedding, which has been shown effective for modeling fluid flows~\cite{arbabiDataDrivenKoopmanModel2018a}.
The observables, or the lifted state, are defined as
\begin{equation}
    \bm{z}_k = \begin{bmatrix} \tilde{\bm{y}}_k^\top, \tilde{\bm{y}}_{k-1}^\top, \ldots, \tilde{\bm{y}}_{k-\tau}^\top, \bm{u}_{k-1}^\top,  \bm{u}_{k-2}^\top, \ldots, \bm{u}_{k-\tau + 1}^\top \end{bmatrix}^\top, \label{eq:observables}
\end{equation}
where $\tilde{\bm{y}} \in \mathbb{R}^{\tilde{n}}$ is the reduced-order velocity field obtained by projecting the full velocity measurement onto a $\tilde{n}$-dimensional subspace using Proper Orthogonal Decomposition (POD), and $\tau$ is the time delay embedding dimension.
POD is used to reduce the dimensionality of the data, improve numerical conditioning, and reduce the computational cost of identifying $\bm{F}$, $\bm{G}$, and $\bm{H}$.
This construction implicitly assumes that the control input affects the system linearly.
However, in our case, the electrode and electromagnet signals $\bm{\phi}$ and $\bm{\psi}$ interact multiplicatively in the Lorentz force term~\eqref{eq:lorentz_force_expanded}.
To address this, we define a virtual input
\begin{equation}
    \bm{u} = \bm{\psi} \otimes \bm{\phi} = \begin{bmatrix}
        \phi_1\psi_1, \phi_2\psi_1, \ldots, \phi_{n_\mathrm{el}}\psi_{n_\text{mag}}
    \end{bmatrix}^\top,
\end{equation}
where $\otimes$ denotes the Kronecker product.

The matrices $\bm{F}$, and $\bm{G}$ are identified using extended dynamic mode decomposition with control (EDMDc) \cite{williamsDataDrivenApproximation2015, kordaLinearPredictorsNonlinear2018}.
Given $D$ samples of the form $(\bm{x}_k, \bm{u}_k, \bm{x}_k^+)$, where $\bm{x}_k$ is the system state, $\bm{u}_k$ is the control input, and $\bm{x}_k^+$ is the successor state, we construct the corresponding data matrices 
\begin{subequations}
    \begin{align}
        \bm{X}_\text{lifted} &= \begin{bmatrix} \bm{z}( \bm{x}_{1} ), \bm{z}( \bm{x}_{2} ), \ldots, \bm{z}( \bm{x}_{D} ) \end{bmatrix}, \\
        \bm{X}^{+}_\text{lifted} &= \begin{bmatrix} \bm{z}( \bm{x}^{+}_{1} ), \bm{z}( \bm{x}^{+}_{2} ), \ldots, \bm{z}( \bm{x}^{+}_{D} ) \end{bmatrix},\\
        \bm{U} &= \begin{bmatrix} \bm{u}_{1}, \bm{u}_{2}, \ldots, \bm{u}_{D} \end{bmatrix}.
    \end{align}
\end{subequations}
The matrices $\bm{F}$, and $\bm{G}$ are obtained by solving the optimization problem
\begin{equation}
    \minimize_{\bm{F}, \bm{G}} \lVert \bm{X}^{+}_\text{lifted} - \bm{F} \bm{X}_\text{lifted} - \bm{G} \bm{U} \rVert_F, \label{eq:edmdc}
\end{equation}
where $\lVert \cdot \rVert_F$ denotes the Frobenius norm. 
This problem has a closed-form solution
\begin{equation}
    \begin{bmatrix} \bm{F} & \bm{G} \end{bmatrix} = \bm{X}^{+}_\text{lifted} \begin{bmatrix} \bm{X}_\text{lifted} \\ \bm{U} \end{bmatrix}^\dagger,
\end{equation}
where $\dagger$ denotes the Moore--Penrose pseudoinverse. 
The convergence of the system \eqref{eq:surrogate_model} to the true Koopman operator is discussed in \cite{kordaConvergenceExtendedDynamic2018b}.

Finally, we construct the output matrix $\bm{H}$. 
Given the structure of our observables \eqref{eq:observables}, $\bm{H}$ extracts the current velocity estimate from the lifted state by projecting it back from the POD subspace to the full measurement space
\begin{equation}
    \bm{H} = \begin{bmatrix}
        \bm{\Xi} & \bm{0}_{\tilde{n} \times \tilde{n}(\tau-1)} & \bm{0}_{\tilde{n} \times n_\mathrm{el}n_\text{mag}(\tau-1)}
    \end{bmatrix},
\end{equation}
where $\bm{\Xi}$ is the POD basis matrix and $\bm{0}_{m \times n}$ denotes the $m \times n$ zero matrix.
    
\subsection*{Koopman model predictive control}
Based on the constructed Koopman operator approximation, we now formulate the optimization problem that the controller solves at each time step.
The controller computes optimal control sequences by solving
\begin{subequations}
    \begin{align}
        \minimize_{\Phi, \Psi} \quad &  J\left(\hat{Y}, \Phi, \Psi\right), \label{eq:mpc_objective_ioopman} \\
        \text{subject to} \quad & \bm{z}_{i+1} = \bm{F}\bm{z}_i + \bm{G}\bm{u}_i, \quad i = 0, 1, \ldots, N_\mathrm{p} - 1, \\
        & \hat{\bm{y}}_i = \bm{H}\bm{z}_i, \\
        & \bm{u}_i = \bm{\psi}_i \otimes \bm{\phi}_i, \label{eq:mpc_input_rank_one_koopman} \\
        & \bm{\phi}_\text{min} \leq \bm{\phi}_i \leq \bm{\phi}_\text{max}, \label{eq:mpc_electrode_constraints_koopman} \\
        & \bm{\psi}_\text{min} \leq \bm{\psi}_i \leq \bm{\psi}_\text{max}, \label{eq:mpc_coil_constraints_koopman} \\
        & \bm{z}_0 = \hat{\bm{z}}_k. 
    \end{align}\label{eq:full_control_mpc_koopman}
\end{subequations}
Here, $N_\mathrm{p}$ is the prediction horizon, $\hat{Y} = \left(\hat{\bm{y}}_i\right)_{i=1}^{\mathrm{N}_p}$  is the predicted output sequence, and $\Phi = \left(\bm{\phi}_i\right)_{i=0}^{\mathrm{N}_p-1}$, $\Psi = \left(\bm{\psi}_i\right)_{i=0}^{\mathrm{N}_p-1}$ are the sequences of electrode and electromagnet control signals, respectively. 
The objective function is defined as
\begin{subequations}
   \begin{align}
    J\left(\hat{Y}, \Phi, \Psi\right) &= \sum_{i=0}^{N_\text{p}-1}  \frac{w_\text{err}}{N_\text{ref}} \lVert \hat{\bm{y}}_{i+1} - \bm{y}^\text{ref}_{i+1}\rVert^2 +  w_\text{el} \lVert \bm{\phi}_i -\bm{\phi}_{i-1} \rVert^2 + w_\text{mag} \lVert \bm{\psi}_i - \bm{\psi}_{i-1}\rVert^2.
\end{align} 
\end{subequations}
where the first term penalizes the deviation from the reference velocity (or vorticity) fields, with the norm taken only over the $N_\text{ref}$ grid points where the reference is specified.
The second and third terms penalize rapid changes in the control signals to ensure smooth actuation. 
The nonnegative weights $w_\text{err}$, $w_\text{el}$, and $w_\text{mag}$ allow us to balance the trade-off between tracking performance and control signal smoothness.
The lifted state $\hat{\bm{z}}_k$ is estimated using a linear Kalman filter (KF), which fuses the current velocity field measurement, control inputs, and model prediction.
The KF is used instead of reconstructing the lifted state \eqref{eq:observables} directly from noisy data to improve robustness. We describe the KF implementation in detail in SI~Appendix~\ref{app:kalman_filter}.
We observed that this significantly improves closed-loop performance in the presence of measurement noise.

When either set of control signals ($\Phi$ or $\Psi$) is fixed in \eqref{eq:full_control_mpc_koopman}, the resulting optimization problem becomes a convex QP in the remaining set.
This property enables an efficient alternating optimization scheme, in which we update $\Phi$ and $\Psi$ in turn until convergence.
We present the complete control loop algorithm with the alternating optimization scheme in Algorithm \ref{alg:mpc_algorithm}.
\begin{algorithm}[tb]
    \begin{algorithmic}[1]
        \For {$k = 0, 1, \ldots$}
            \State Measure velocity field $\bm{y}_k$ and estimate $\hat{\bm{z}}_k$ using the Kalman filter
            \State Initialize $\Phi^{(0)}$ (e.g., randomly, or using the previous solution)
            \State Set $n = 0$
            \Repeat
                \State Solve \eqref{eq:full_control_mpc_koopman} for $\Psi^{(n+1)}$ while fixing $\Phi^{(n)}$
                \State Solve \eqref{eq:full_control_mpc_koopman} for $\Phi^{(n+1)}$ while fixing $\Psi^{(n+1)}$
                \State $n \gets n + 1$
            \Until{$\Phi^{(n)} \approx \Phi^{(n-1)}$}
            \State Set $\Phi^\star = \Phi^{(n)}$ and $\Psi^\star = \Psi^{(n)}$
            \State Apply the signals $\bm{\phi}_k = \bm{\phi}_{0}^\star$ and $\bm{\psi}_k =  \bm{\psi}_{0}^\star$ to the electrodes and electromagnets
        \EndFor
    \end{algorithmic}
    \caption{Koopman MPC control loop with alternating optimization}
    \label{alg:mpc_algorithm}
\end{algorithm}

\subsection*{Data collection}
The Koopman-based model was trained on data generated from three-dimensional FEM simulations of the incompressible Navier--Stokes equations
\begin{subequations}
    \begin{align}
        \frac{\partial \bm{v}}{\partial t} + \left(\bm{v} \cdot \nabla \right) \bm{v} &= -\nabla p + \nu \nabla^2 \bm{v} + \frac{1}{\rho} \bm{f},\\
        \nabla \cdot \bm{v} &= 0,
    \end{align} \label{eq:navier_stokes}
\end{subequations}
where $\bm{v}$ is the velocity field, $p$ is the pressure (normalized by the fluid density), $\nu$ is the kinematic viscosity, $\rho$ is the fluid density, and $\bm{f}$ is the Lorentz force density, which is given by \eqref{eq:lorentz_force_expanded}.
The simulations were performed over a cylindrical domain of the experimental dish, with the no-slip boundary condition applied at the walls and a slip condition at the fluid surface.
Electric and magnetic fields were constructed by superposing precomputed spatial bases for each actuator, obtained via high-fidelity FEM simulations of the full platform in COMSOL, and scaled by time-varying control signals as per \eqref{eq:fields_decomposition}.
Employed numerical methods and mesh details are described in SI~Appendix~\ref{app:simulation}.

Actuator signals were generated by a randomized signal generator that either held the previous value or sampled a new one from a uniform distribution over actuator-specific bounds.
This design ensured persistent excitation while maintaining physical plausibility; see SI~Appendix~\ref{app:identification_signal} for details.

The dataset comprised 7 trajectories of 20{,}000 time steps each at \SI{0.5}{\second} intervals.
Using the system's geometric symmetries---rotations and reflections---we augmented the dataset to 56 trajectories.
SI~Appendix~\ref{app:symmetries} describes the used symmetries in detail.

\subsection*{Experimental setup}
\begin{figure}[tbhp]
    \centering
    \includegraphics[width=\textwidth]{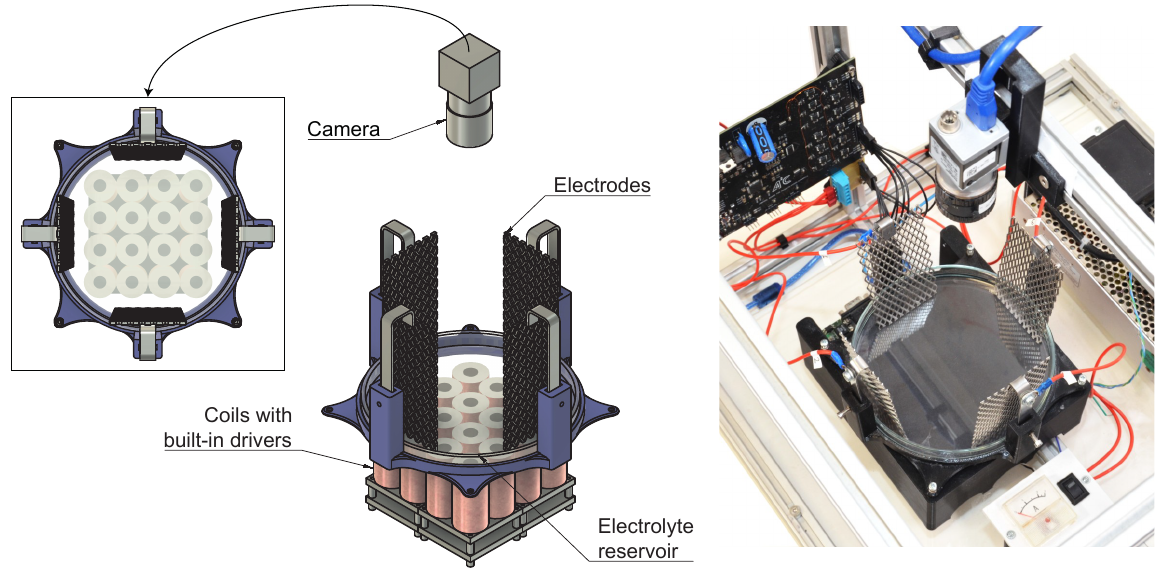}
    \caption{
Experimental platform used for flow control experiments. The system consists of a shallow cylindrical reservoir filled with conductive electrolyte, actuated by four electrodes placed around the rim and a stack of solenoidal coils with integrated drivers. A top-mounted camera records tracer particle motion for velocity reconstruction. \textbf{Left:} CAD model showing components and geometry. \textbf{Right:} Photograph of the assembled experimental setup.
}
    \label{fig:platform-real}
\end{figure}

An overview of the experimental setup is shown in Figure~\ref{fig:platform-real}. 
It consists of a shallow dish filled with a diluted sulfuric acid solution, with a measured electrical conductivity of $\sigma = \SI{5}{\siemens\per\metre}$.
The dish has a diameter of \SI{143}{\milli\metre}, and the fluid depth is \SI{8}{\milli\metre}.

Four mesh-pattern titanium electrodes coated with platinum are placed around the periphery of the dish. 
Each electrode is connected to a custom-built controllable power supply capable of applying voltages in the range of $0$–$\SI{10}{\volt}$ to each electrode independently. 
The power supply is interfaced with a computer via USB.
The dish is mounted on a custom-built platform housing 16 solenoidal electromagnets arranged in a $4 \times 4$ grid. 
The electromagnets are grouped into modules, each consisting of four coils and their corresponding driver boards. 
The modules communicate with the control computer using the RS-485 protocol, and allow bidirectional current control in the range of $\pm\SI{440}{\milli\ampere}$. 
For simplicity, only the four central electromagnets are used in the experiments presented.

The surface of the fluid is seeded with \SI{85}{\micro\metre} polymethylmethacrylate (PMMA) tracer particles. 
Particle motion is recorded by a 1-megapixel grayscale camera positioned above the dish, capturing a \SI{10}{\centi\metre} $\times$ \SI{10}{\centi\metre} region at \SI{25}{\hertz}. 
The camera connects via USB to the control computer, which runs a real-time PIV algorithm to reconstruct the planar velocity field. 
The algorithm uses interrogation windows of $32 \times 32$ pixels with \SI{50}{\percent} overlap and parabolic subpixel interpolation.
While electrolysis at the electrodes occasionally produced gas bubbles, they remained outside the field of view and did not interfere with PIV measurements.

\newpage

\subsection*{Parameters}
The parameters used in the experiments are summarized in Table~\ref{tab:params}.
The same values were used across all reference tracking experiments presented in the Results section.

\begin{table}[h]
\centering
\caption{Parameter values used in the experiments.}
\begin{tabular}{ll}
\toprule
\textbf{Parameter} & \textbf{Value} \\
\midrule
Electrical conductivity ($\sigma$) & \SI{5}{\siemens\per\metre} \\
Number of electrodes ($n_\text{el}$) & 4 \\
Number of electromagnets ($n_\text{mag}$) & 4 \\
POD subspace dimension ($\tilde{n}$) & 512 \\
Time delay embedding dimension ($\tau$) & 4 \\
Time step ($\Delta t$) & \SI{0.5}{\second} \\
Prediction horizon ($N_\mathrm{p}$) & 10 (\SI{5}{\second}) \\
Tracking error weight ($w_\text{err}$) & $10^4$ \\
Electrode weights ($w_\text{el}$) & $10^{-2}$  \\
Electromagnets weight ($w_\text{mag}$) & $10^{0}$  \\
\bottomrule
\end{tabular}
\label{tab:params}
\end{table}

\FloatBarrier


\bibliographystyle{unsrt}
\bibliography{data-driven-control-of-a-magnetohydrodynamic-flow}

\newpage

\appendix

\pagenumbering{arabic}      
\setcounter{page}{1}        
\renewcommand{\thepage}{S\arabic{page}}  

\section*{SI Appendix}
\renewcommand{\thesubsection}{S\arabic{subsection}}
\setcounter{figure}{0}
\renewcommand{\thefigure}{S\arabic{figure}}

\renewcommand{\thetable}{S\arabic{table}}
\setcounter{table}{0}

\renewcommand{\theequation}{S\arabic{equation}}
\setcounter{equation}{0}

\subsection{Derivation and assumptions of the Lorentz force model}\label{app:lorentz_force}
The Lorentz force acting on the fluid is given as
\begin{equation}
    \bm{f} = \bm{J} \times \bm{B}, 
\end{equation}
where $\bm{J}$ is the current density in the fluid, and $\bm{B}$ is the magnetic field.
Our first assumption is that we work in the low magnetic Reynolds number regime so we can neglect the magnetic field generated by the current density $\bm{J}$ in the fluid, and thus $\bm{B}$ is completely externally generated.
This is a well-justified assumption for flows of electrolytes \cite{davidsonIntroductionMagnetohydrodynamics2016}.
The current density is described by Ohm's law
\begin{equation}
    \bm{J} = \sigma \left( \bm{E} + \bm{v} \times \bm{B} \right),
\end{equation}
where $\sigma$ is the conductivity of the fluid, $\bm{E}$ is the electric field, $\bm{v}$ is the velocity of the fluid.
Our second assumption is that $\lVert \bm{E} \rVert \gg \lVert \bm{v} \times \bm{B} \rVert$, therefore the second term in the Ohm's law can be neglected, and the current density simplifies to
\begin{equation}
    \bm{J} = \sigma \bm{E}.
\end{equation}
This assumption holds in our experimental setup, where $\lVert \bm{E} \rVert \sim \SI{1}{\volt\per\metre}$ dominates $\lVert \bm{v} \times \bm{B} \rVert \sim \SI{10}{\micro\volt\per\metre}$, yielding a ratio of $\lVert \bm{v} \times \bm{B} \rVert / \lVert \bm{E} \rVert \sim 10^{-5}$.
The Lorentz force can be then rewritten as
\begin{equation}
    \bm{f} = \sigma \bm{E} \times \bm{B}. \label{eq:lorentz_force_simplified}
\end{equation}
Now, we assume that the electric and magnetic fields are generated by superposing the fields generated by $n_\mathrm{el}$ electrodes and $n_\text{mag}$ electromagnets, respectively, while the field of each actuator is decomposable into a time-varying control signal and a spatially-varying field, i.e.,
\begin{equation}
    \bm{E}(t, \bm{r}) = \sum_{i=1}^{n_\mathrm{el}} \phi_i(t)\bm{E}_{i}(\bm{r}), \quad \bm{B}(t, \bm{r})  = \sum_{j=1}^{n_\text{mag}} \psi_j(t) \bm{B}_{j}(\bm{r}),
\end{equation}
where $\bm{E}_{i}(\bm{r})$ and $\bm{B}_{j}(\bm{r})$ are the spatially-varying components of the electric and magnetic fields generated by the $i$-th electrode and the $j$-th electromagnet, respectively, while $\phi_i(t)$ and $\psi_j(t)$ are the time-varying control signals applied to the respective electrodes and electromagnets.
The Lorentz force density \eqref{eq:lorentz_force_simplified} can then be expressed as
\begin{equation}
    \bm{f}(t, \bm{r}) = \sigma \sum_{i=1}^{n_\mathrm{el}} \sum_{j=1}^{n_\text{mag}} \phi_i(t)\psi_j(t) \bm{E}_{i}(\bm{r}) \times  \bm{B}_{j}(\bm{r}). \label{eq:app_lorentz_force_expanded}
\end{equation}

\subsection{Electric and magnetic fields decomposition}\label{app:fields_decomposition}
Each of the spatial components of the electric fields is given as $\bm{E} = -\nabla V$, where $V$ is the electric potential---the solution of the Laplace equation $\nabla^2 V = 0$ with boundary conditions where $V$ equals $V_\text{ref}$ on the surface of the given electrode, while it equals zero on the other electrodes.
Similarly, each of the spatial components of the magnetic fields is given as $\bm{B} = \nabla \times \bm{A}$, where $\bm{A}$ is the magnetic vector potential---the solution of $\nabla \times \nabla \times \bm{A} = \mu \bm{J}_\text{coil}$ with $\mu$ being the magnetic permeability of the environment, and $\bm{J}_\text{coil}$ the coil current density.
We adopt the homogenized multiturn model for the coil current density, where $\bm{J}_\text{coil} = N I_\text{ref} / A \bm{e}$, with $N$ being the number of turns, $I_\text{ref}$ the reference coil current, while $A$ is the area and $\bm{e}$ the normal vector to the cross-section of the coil's winding. 

We select $V_\text{ref} = \SI{1}{\volt}$ as the reference voltage applied to the electrodes. The control signal $\phi(t)$ then represents the ratio of the applied voltage to the reference voltage, and is bounded between $0$ an $10$, as our setup allows for voltages in the range $0$–$\SI{10}{\volt}$.

Coils usually contain a ferromagnetic core, which significantly increases the magnetic field strength, but also makes the magnetic field nonlinear with respect to the current due to the saturation effect.
This can be accounted for by using nonlinear relationship between the control signal and the respective coil current, i.e., instead of defining the control signal $\psi$ as the ratio of the coil current to a reference current $\psi = I_\text{coil}/I_{\text{ref}}$, we employ a nonlinear transformation $f$ such that $\psi = f(I_\text{coil}/I_{\text{ref}})$.
For our setup, we found $f$ by measuring the magnetic field strength as a function of the coil current and fitting a nonlinear function to the data.
Specifically, we use
\begin{equation}
    f(x) = 0.889 \arctan{\left( 1.104 \left| x \right| + 12.227\left| x \right|^2 \right)} \sign\left(x\right),
\end{equation}
with $I_{\text{ref}} = \SI{440}{\milli\ampere}$. Importantly, $f$ is invertible, enabling us to recover the coil current from the signal. This allows the optimization problem to be formulated in terms of control signals, while still accounting for the nonlinear relationship between the coil current and the magnetic field strength.
Furthermore, by this construction, the control signal $\psi$ is bounded between $-1$ and $1$, as the coils allow for bidirectional current control.

\subsection{Solution of the optimization problem}
We now show how one-step of the alternating optimization scheme is solved.
Specifically, the alternating optimization scheme is composed of the QP subproblems of the form
\begin{subequations}
    \begin{align}
        \minimize_{\nu} \quad & \frac{1}{2} \sum_{i=0}^{N_\text{p}-1} \frac{w_\text{err}}{N_\text{ref}}  \lVert \hat{\bm{y}}_{i+1} -  \bm{y}^\text{ref}_{i+1} \rVert^2 +  w \lVert \bm{\nu}_{i} -  \bm{\nu}_{i-1} \rVert^2, \\
        \text{subject to} \quad & \bm{z}_{i+1} = \bm{F}\bm{z}_i + \bm{G}_{i}\bm{\nu}_{i}, \quad i = 0, 1, \ldots, N_\mathrm{p} - 1, \label{eq:predictor_constraint} \\ 
        & \hat{\bm{y}}_i = \bm{H}\bm{z}_i, \quad i = 1, 2, \ldots, N_\mathrm{p},\\
        & \bm{\nu}_{\text{min}} \leq \bm{\nu}_{i} \leq \bm{\nu}_{\text{max}}, \quad i = 0, 1, \ldots, N_\mathrm{p} - 1,\\
        & \bm{z}_0 = \hat{\bm{z}}_k,
    \end{align}
    \label{eq:alternating_optimization}
\end{subequations}
where $\nu \in \{\phi, \psi\}$, and $w \in \{w_\text{el}, w_\text{mag}\}$, i.e., the formulation covers both cases of fixing the electrode signals and optimizing over the coil signals and vice versa, while $N_\text{ref}$ is the number of grid points where the reference velocity field is specified.
Furthermore, the problems feature time-varying matrices $\bm{G}_{i}$ in the predictor constraint \eqref{eq:predictor_constraint}.
These time-varying input matrices are constructed from the other set of commands, than the one being optimized, and the constant input matrix $\bm{G}$ of the Koopman model as
\begin{equation}
    \bm{G}_{i} = \bm{G}\left(\bm{\phi}_i \otimes \bm{I}_{n_\text{mag}\times n_\text{mag}}\right),
\end{equation}
in case of fixing the electrode signals and optimization over the coil signals, and
\begin{equation}
    \bm{G}_{i} = \bm{G}\left(\bm{I}_{n_\text{el}\times n_\text{el}} \otimes\bm{\psi}_i\right), 
\end{equation}
in the other case. Here, $\otimes$ denotes the Kronecker product, and $\bm{I}_n$ is the identity matrix of size $n\times n$.

We can reformulate these subproblems into so-called dense formulation
\begin{subequations}
    \begin{align}
        \minimize_{\bar{\bm{\nu}}} \quad & \frac{1}{2} \bar{\bm{\nu}}^\top \bm{P} \bar{\bm{\nu}} +  \bm{q}^\top \bar{\bm{\nu}},\label{eq:qp_objective}\\
        \text{subject to} \quad & \bar{\bm{\nu}}_{\text{min}} \leq \bar{\bm{\nu}}\leq \bar{\bm{\nu}}_{\text{max}}, \label{eq:qp_constraints}
    \end{align}
\end{subequations}
where all variables but the control signals are eliminated, and the sequence of control signals is stacked into a single vector $\bar{\bm{\nu}}$.
This results in a QP that is releatively small because we expect the number of electrodes and coils to be much smaller than the dimension of the lifted state.
We now detail the construction of the hessian $\bm{P}$ and the linear term $\bm{q}$.
The hessian takes the form
\begin{equation}
   \bm{P} = \frac{w_\text{err}}{N_\text{ref}}\bar{\bm{G}}^\top\bar{\bm{H}}^\top\bar{\bm{H}}\bar{\bm{G}} + w \bm{D},
\end{equation}
where 
\begin{equation}
\bar{\bm{G}} = 
\begin{bmatrix}
    \bm{G}_{1} & \bm{0} & \ldots & \bm{0}\\
    \bm{F}\bm{G}_{1} & \bm{G}_{2} & \ldots & \bm{0}\\
    \vdots & \vdots & \ddots & \vdots\\
    \bm{F}^{N_\mathrm{p}-1}\bm{G}_{1} & \bm{F}^{N_\mathrm{p}-2}\bm{G}_{2} & \ldots & \bm{G}_{N_\mathrm{p}}
\end{bmatrix}, \quad \bar{\bm{H}} = \bm{I}_{N_\mathrm{p}} \otimes \bm{H},
\end{equation} 
and
\begin{equation}
    \bm{D} =
    \begin{bmatrix}
        2 & -1 & 0 & \cdots & 0 & 0 \\
        -1 & 2 & -1 & \cdots & 0 & 0 \\
        0 & -1 & 2 & \cdots & 0 & 0 \\
        \vdots & \ddots & \ddots & \ddots & \ddots & \vdots \\
        0 & 0 & \cdots & -1 & 2 & -1 \\
        0 & 0 & \cdots & 0 & -1 & 1 \\
        \end{bmatrix} \otimes \bm{I},
\end{equation}
where $\bm{I}$ is the identity matrix of either size $n_\mathrm{el}\times n_\mathrm{el}$ or $n_\text{mag}\times n_\text{mag}$, depending on optimized control signals.
The linear term is constructed as
\begin{equation}
    \bm{q} = \bm{W}^\top\hat{\bm{z}}_k - \bm{S}^\top\bm{\nu}_0 - \bm{T}^\top\begin{bmatrix} (\bm{y}_1^\text{ref})^\top & \cdots & (\bm{y}_{N_\mathrm{p}}^\text{ref})^\top \end{bmatrix}^\top,
\end{equation}
where 
\begin{equation}
    \bm{W} = w_\text{err}\bar{\bm{F}}^\top\bar{\bm{H}}^\top\bar{\bm{H}}\bar{\bm{G}},\quad \bm{S} = \begin{bmatrix} w\bm{I}_n & \bm{0}_{\left(N_\mathrm{p}-1\right)n\times n}     \end{bmatrix}, \quad \bm{T} = w_\text{err}\bar{\bm{H}}^\top\bar{\bm{H}}\bar{\bm{G}},
\end{equation}
and 
\begin{equation}
    \bar{\bm{F}} = \begin{bmatrix}
        \bm{F}\\
        \bm{F}^2\\
        \vdots\\
        \bm{F}^{N_\mathrm{p}}
    \end{bmatrix}.
\end{equation}

\subsection{Convergence of the alternating optimization scheme}\label{app:convergence}
We solve the non-convex MPC problem by alternately minimizing over the electrode and electromagnet signals.
Because each QP subproblem has a positive-definite Hessian, it is strongly convex with a unique minimizer. Established convergence theory for block-coordinate methods (e.g. \citeapp{tsengConvergenceBlockCoordinate2001}) then guarantees the following.
\begin{enumerate}
    \item Each update never increases the MPC cost.
    \item The iterates settle to a solution where neither the electrode nor the electromagnet signals can be adjusted alone to further lower the cost. 
\end{enumerate}
The resulting control signals form a coordinate-wise optimum, meaning they satisfy the first-order optimality conditions for each block of variables. 
This critical point may be then a local minimum or a saddle point of the full MPC problem.   

\subsection{Kalman filter}\label{app:kalman_filter}
We estimate the lifted state $\hat{\bm{z}}_k$ from the velocity field measurements $\bm{y}_i$, $i = 0, 1, \ldots, k$, and the virtual control inputs $\bm{u}_i$, $i = 0, 1, \ldots, k-1$, using a linear Kalman filter (KF).  
The filter assumes that the lifted system dynamics follow
\begin{align}
    \bm{z}_{k+1} &= \bm{F}\bm{z}_k + \bm{G}\bm{u}_k + \bm{w}_k,\\
    \bm{y}_k &= \bm{H}\bm{z}_k + \bm{v}_k,
\end{align}
where $\bm{w}_k \sim \mathcal{N}(\bm{0}, \bm{Q})$ and $\bm{v}_k \sim \mathcal{N}(\bm{0}, \bm{R})$ represent the process and measurement noise, respectively, with covariance matrices $\bm{Q}$ and $\bm{R}$.
The Kalman filter proceeds recursively via
\paragraph*{Prediction step}
\begin{align}
    \hat{\bm{z}}_{k|k-1} &= \bm{F} \hat{\bm{z}}_{k-1|k-1} + \bm{G} \bm{u}_{k-1}, \\
    \bm{P}_{k|k-1} &= \bm{F} \bm{P}_{k-1|k-1} \bm{F}^\top + \bm{Q},
\end{align}
\paragraph*{Update step}
\begin{align}
    \bm{K}_k &= \bm{P}_{k|k-1} \bm{H}^\top \left( \bm{H} \bm{P}_{k|k-1} \bm{H}^\top + \bm{R} \right)^{-1}, \\
    \hat{\bm{z}}_{k|k} &= \hat{\bm{z}}_{k|k-1} + \bm{K}_k \left( \bm{y}_k - \bm{H} \hat{\bm{z}}_{k|k-1} \right), \\
    \bm{P}_{k|k} &= \left( \bm{I} - \bm{K}_k \bm{H} \right) \bm{P}_{k|k-1},
\end{align}
where $\hat{\bm{z}}_{k|k}$ is the state estimate at time $k$ given measurements up to $k$, and $\bm{P}_{k|k}$ is the corresponding estimation error covariance.  
The filter is initialized with $\hat{\bm{z}}_{0|0} = \bm{0}$ and $\bm{P}_{0|0} = \bm{0}$.

We found the process noise covariance $\bm{Q}$ and measurement noise covariance $\bm{R}$ heuristically by tuning them to achieve a good trade-off between responsiveness and noise suppression.  
With $\bm{Q}$, we mirror the delay embedding structure of the Koopman model, assuming that the process noise affects only the delayed velocity measurements and not the delayed inputs.  
This yields the block structure
\begin{equation}
    \bm{Q} = \begin{bmatrix}
        q \bm{I}_{\tilde{n}\tau} & \bm{0} \\
        \bm{0} & \bm{0}
    \end{bmatrix},
\end{equation}
where $\tilde{n}$ is the number of POD modes, $\tau$ is the number of delay steps, and $n_\text{el}, n_\text{mag}$ are the number of electrode and magnet inputs, respectively.  
The total lifted state dimension is $\tilde{n}\tau + n_\text{el}n_\text{mag}(\tau - 1)$.
The measurement noise covariance is
\begin{equation}
    \bm{R} = r \bm{I}_{2n_x n_y},
\end{equation}
where $n_x = 32$ and $n_y = 32$ are the number of grid points in the $x$ and $y$ directions of the PIV measured velocity field.
We selected $q = 10^{-8}$ and $r = 5 \times 10^{-5}$.

\subsection{Identification signal}\label{app:identification_signal}
We generate the identification signal for identifing the Koopman-based model using the following procedure.
First, we generate a random signal sequence $\bm{\nu}_1, \bm{\nu}_2, \ldots, \bm{\nu}_N$, where $N$ is the length of the identification signal, and $\nu \in \{\phi, \psi\}$.
The sequence is generated by the following generator
\begin{equation}
    \bm{\nu}_k = \begin{cases}
        \bm{\nu}_{k-1} & \text{if } q_k \leq p,\\
        \bm{s}_{k} & \text{if } q_k > p,
    \end{cases}
\end{equation}
where $\bm{s}_{k} \sim \mathcal{U}{\left(\left[\bm{\nu}_\text{min}, \bm{\nu}_\text{max} \right]\right)}$ is a random signal drawn from a vector-valued uniform distribution, $p$ is a probability threshold, and $q_k \sim \mathcal{U}{\left(\left[ 0, 1 \right]\right)}$ is a random number.
We employed the rule $p = 1-\Delta t / \tau$, where $\Delta t$ is the control loop timestep, and $\tau$ is the desired mean switching time of the signals, in our experiments $\tau = \SI{80}{\second}$.
The sequence is then filtered using a forward-backward pass of a first order Butterworth filter with a cutoff frequency of $\SI{0.1}{\hertz}$ to reduce high-frequency components.

\subsection{Simulation}\label{app:simulation}
\begin{figure}[tb]
    \includegraphics[width=\textwidth]{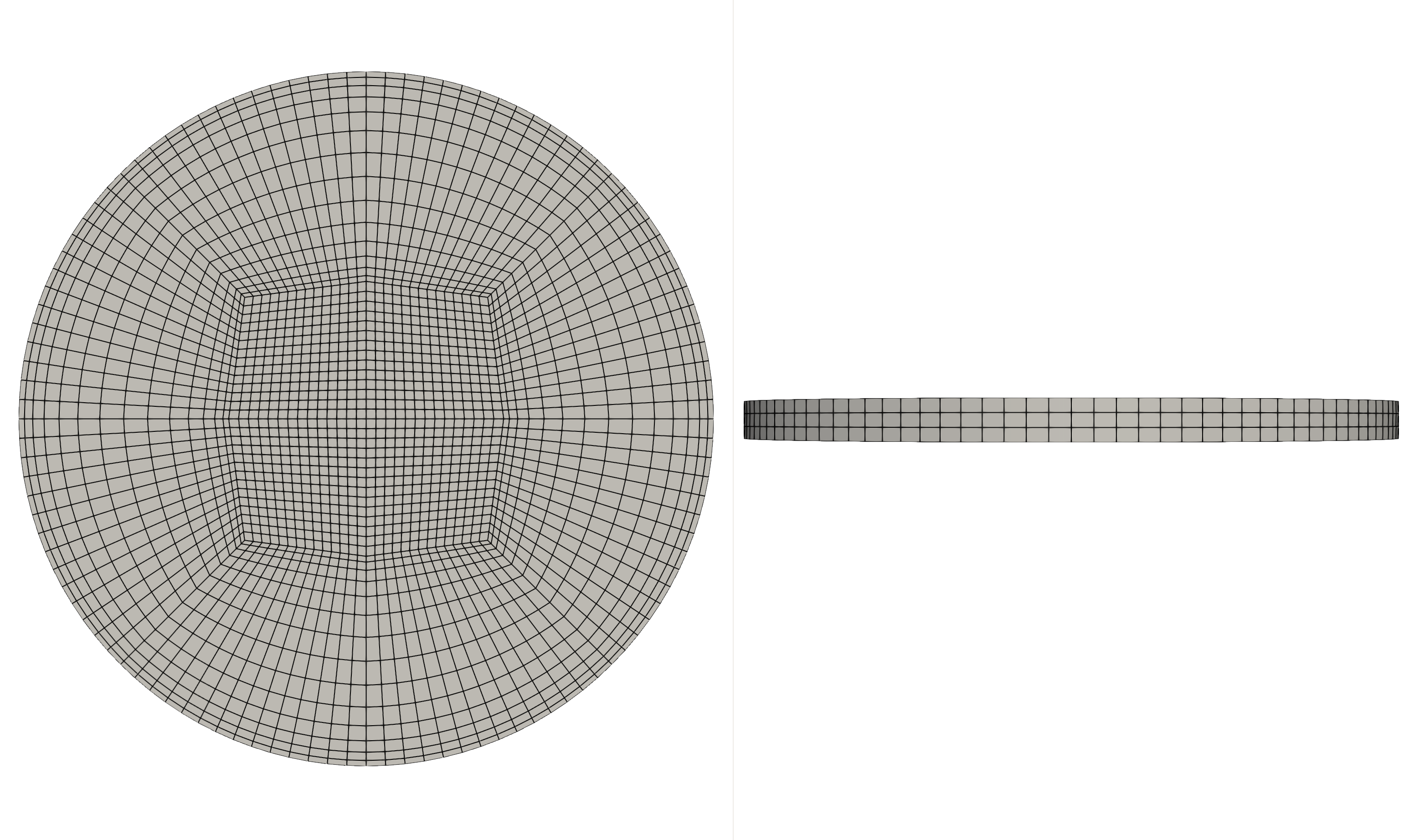}
    \caption{Structured hexahedral mesh used in the simulation of the cylindrical fluid domain (diameter \SI{143}{\milli\metre}, height \SI{8}{\milli\metre}).
Left: top-down (XY) view showing radial, and central refinement.
Right: side (XZ) view showing uniform vertical discretization.
    }
    \label{fig:mesh} 
\end{figure}

The training data for the Koopman-based model was generated using a custom finite element simulation of the incompressible Navier--Stokes equations over a cylindrical domain matching the experimental dish (diameter \SI{143}{\milli\metre}, height \SI{8}{\milli\metre}). 
We used the Ferrite.jl package in Julia with structured hexahedral Q2/Q1 elements (quadratic velocity, linear pressure), and we show the mesh used in Figure \ref{fig:mesh}.
Time integration employed a semi-implicit BDF2 scheme with extrapolated convective term, and the velocity-pressure coupling was handled using the Incremental Pressure Correction Scheme (IPCS) in rotational form. 
The time step was fixed at $\Delta t = \SI{0.25}{\second}$, twice the experimental sampling rate, to ensure stability.
Boundary conditions were no-slip on the bottom and sidewalls, and free-slip on the top surface.

Lorentz forces were applied as time-varying body forces using the bilinear model \eqref{eq:app_lorentz_force_expanded}, with the basis fields $\bm{E}_i(\bm{r})$ and $\bm{B}_j(\bm{r})$ generated in COMSOL.
These fields were exported from COMSOL as values of the Gauss points of the COMSOL mesh and interpolated to the simulation mesh using nearest-neighbor interpolation.

\subsection{Symmetries used for dataset augmentation}\label{app:symmetries}
\begin{figure}[H]
    \centering
    \begin{subfigure}{0.48\textwidth}
        \centering
        \includegraphics[width=0.90\textwidth]{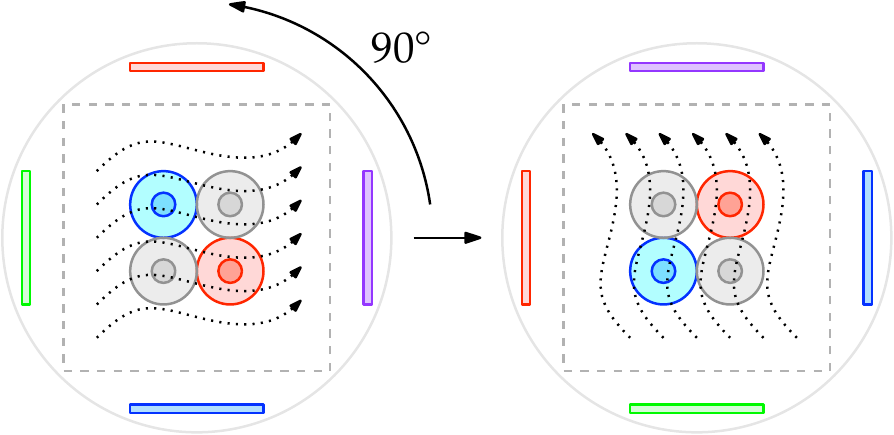} 
        \caption{Rotational symmetry.}
    \end{subfigure}
    \begin{subfigure}{0.48\textwidth}
        \centering
        \includegraphics[width=0.90\textwidth]{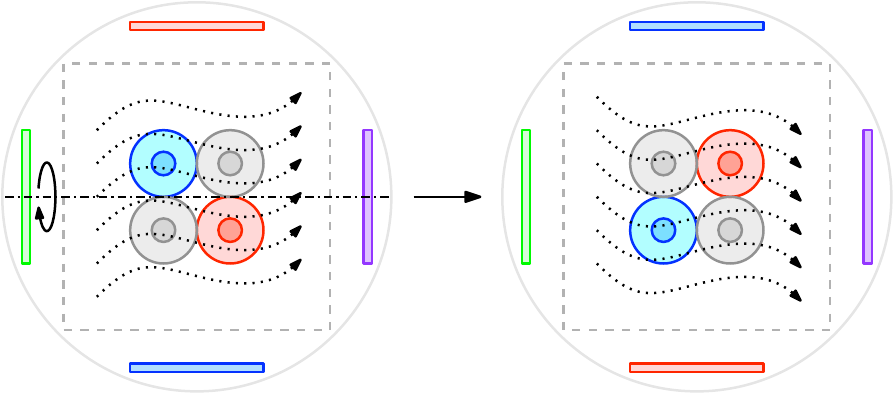}
        \caption{Symmetry about the $x$ axis.}
    \end{subfigure}
    \begin{subfigure}{0.48\textwidth}
        \centering
        \vspace{0.5cm}
        \includegraphics[width=0.90\textwidth]{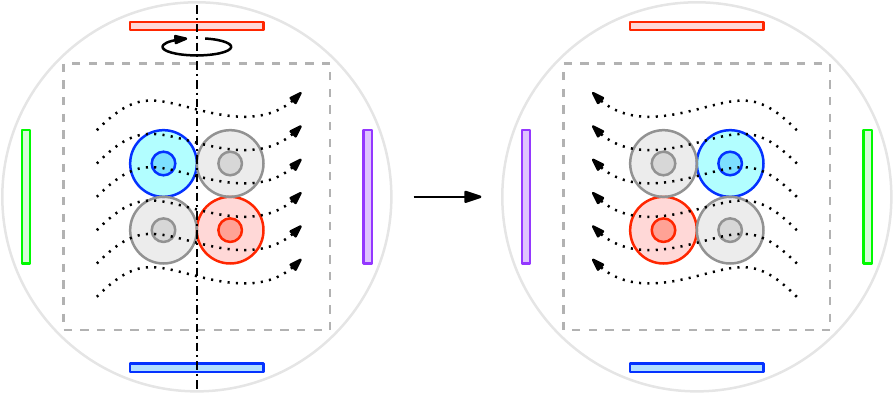}
        \caption{Symmetry about the $y$ axis.}
    \end{subfigure}
    \caption{Symmetries used to augment the dataset.}
    \label{fig:symmetries}
\end{figure}

We exploit the symmetries of the system to augment the dataset.
We show the symmetries used in Figure \ref{fig:symmetries}.

\bibliographystyleapp{unsrt}
\bibliographyapp{si-appendix}

\end{document}